\documentstyle[twocolumn]{mn}

\title{ Local Axisymmetric Instability Criterion in the Thin, Rotating, 
Multicomponent Disk.}

\author[R.R.~Rafikov]
       {R.R.~Rafikov$^1$ \\
           $^1$ Peyton Hall, Princeton University, Princeton, NJ, 08544, USA}
\date{Accepted 2000  .
     Received  2000 ;
     in original form 2000}

\pagerange{\pageref{firstpage}--\pageref{lastpage}}

\pubyear{2000}

\begin{document}

\maketitle

\label{firstpage}

\begin{abstract}
Purely gravitational perturbations are considered in a thin
rotating disk composed of several 
 gas and stellar components.
The dispersion relation for the axisymmetric 
density waves propagating through
the disk is found and the criterion for the local axisymmetric stability of
the whole system is formulated. In the appropriate limit of two-component gas 
we confirm the findings of Jog \& Solomon (1984) and extend consideration
to the case when one component is collisionless. Gravitational 
stability of the Galactic disk in the  Solar neighborhood 
based on the multicomponent instability condition is explored using
recent measurements of the stellar composition and 
kinematics in the local Galactic disk obtained by $Hipparcos$ satellite.
\end{abstract}


\section{Introduction}\label{intro}

It was first shown by Safronov 
(1960) that a thin gaseous rotating disk
 can become unstable against local axisymmetric
 perturbations under the action of its own gravity.
 Quantitatively
the instability criterion is expressed in terms of 
Toomre's $Q$:
\begin{equation}
Q\equiv\frac{\kappa c}{\pi G \Sigma},
\label{Q}
\end{equation}
with $\kappa$ being the epicycle frequency, $c$ being the speed of sound,
and $\Sigma$ being the surface density of the disk. Instability
arises when $Q<1$.

Disks composed of stars behave very similarly to the gaseous disks
for long wavelength because epicyclic motion of stars is the same as
that of the gas. But collisionless disks of stars
have no pressure to resist short wavelength density perturbations and thus
the density waves in this case are different from simple sound waves
present in the gaseous disks in this regime (Toomre 1964).
When the
wavelength becomes shorter than  the typical amplitude of the
star's epicyclic  motion the gravitational effect of the perturbation
averages out and stars perform almost pure epicyclic motion with
 frequency $\kappa$. Instability can occur
though for some intermediate wavelength as in the case of the gaseous
disks.
Toomre (1964) had shown that in the stellar disks self-gravity can support
exponentially growing axisymmetric modes when the condition $Q<1$ is
satisfied
with $\pi$ replaced by  $3.36$ and $c$ replaced by
the stellar velocity dispersion $\sigma$ in the radial direction  in
definition (\ref{Q}).
Except for the qualitative difference in behavior at small wavelength,
stellar and gaseous disks behave almost identically for large wavelength
which leads to a close similarity of the corresponding instability
conditions.

However, real disks of spiral galaxies do not  consist of only one
component
which could be characterized by a single value of the velocity dispersion.
Indeed, a 
significant part  of the ISM in galactic disks is usually concentrated
in the form of cold gas with characteristic sound velocity $\sim 5-10$
km s$^{-1}$. At the same time stars usually have larger velocity dispersions
$\sim 20-35$ km s$^{-1}$.

The problem of the treatment of such
two-component disks was first addressed by
Jog \& Solomon (1984a, JS). They assumed galactic disk to consist of two
fluid
components:
one hot, playing the role of the stellar population, and one cold
representing the usual ISM, and they  neglected the fact that in
 real galaxies hot component is
collisionless and provides no damping. Though this could be
important shortcoming of the analysis for the short wavelength
it turns out to be a good approximation for the stability purposes
as we will see later in this paper.
Jog \& Solomon (JS) have demonstrated that
 even relatively small
amounts of the cold gas ($\sim 10\%$ by mass)
can very effectively destabilize the whole system.
 
The two-fluid stability criterion was applied to a variety of
disks by Jog \& Solomon (1984b), Elmegreen (1995), and Jog (1996).
They showed that our Galaxy is stable against local
axisymmetric perturbations to the extent of the accuracy
of the parameters and mass distribution models assumed.
Theory can be further refined by 
 including
the effects of the finite thickness of the disk on its stability
(Toomre 1964)
and by taking into account the vertical motions in the disk (Romeo 1992).
It was shown that
the instability condition remains the same provided the unperturbed
 surface density is multiplied by an appropriate reduction factor.

Real galaxies always contain more than two isothermal components. 
It is   now widely believed that ISM in the galactic disk is subdivided 
into a number of components with different temperatures and, thus, 
different dynamical properties (McKee \& Ostriker 1977).
 The same is true about the stellar population.
Stars in the galactic disk are being constantly scattered by giant
molecular clouds 
and transient spiral arms which steadily increase their velocity
dispersions (Spitzer \& Schwarzschild 1951, 1953; Barbanis \& Woltjer 1967;
Carlberg \& Sellwood 1985). 
It means that the stars of different ages have different 
velocity dispersions and thus dynamically should be treated separately. 

Recent results of the $Hipparcos$ satellite (ESA 1997) provided us with a
wealth of information about the local stellar kinematics.
The proper motions of different star populations were 
accurately measured and the resulting velocity ellipsoids  
were constructed for them  enabling one 
to study different kinematic constituents
of our Galaxy  separately (Dehnen \& Binney 1998; Mignard 2000). 
This mission also
gave us a lot of information about distances to the stars
which provided us with the local densities of various types of the stars
(Holmberg \& Flynn 2000).

In this paper we present the analysis of the 
axisymmetric gravitational stability of the thin rotating disk
which consists of a number of components, each of them 
being characterized by its 
own temperature and surface density. 
In doing so we distinguish between the two types of components:
collisional and thus having pressure forces such as  usual gas, or
collisionless such as stellar component, which needs a kinetic 
treatment. The derivation of the dispersion relation is presented
in \S \ref{eqns} and \ref{disp-rel}.
In \S \ref{2-fluid} we compare our stability condition with the one 
derived in JS to test the difference arising
when one of the disk components is treated as collisionless which 
better represents real galaxies. Finally, in \S   \ref{appl}
we apply our results to the case of our Galaxy in the Solar neighborhood 
using recent data from $Hipparcos$ satellite.

\section{Basic equations}\label{eqns}

We work in non-rotating cylindrical system, $r,\phi,z$, such that 
$z$-axis coincides with the rotation axis of the disk and angle $\phi$
increases in the direction of rotation. 
We start first with the equations for the fluid components and
then make some correction to  take collisionless components into account.
We will usually use index $i$ to describe  gas components and index
$j$ for stellar components.

All the components are spatially concentrated in the thin 
disk and the effects of the finite disk thickness will usually be
disregarded. All the motions are assumed to occur
only in the plane of the disk. 

We suppose that there are $n_g$ gaseous components contributing
to the mass of the Galaxy, each characterized by the sound velocity
$c_i$ and surface density $\Sigma_{gi}$, and $n_s$ collisionless
components with $\Sigma_{s j}$ being the surface density of 
the $j$-th component. Each collisionless component in the unperturbed state 
 is assumed to
have a Schwarzschild distribution function, that is for $j$-th
component
\begin{eqnarray}
f(v_r,v_\phi)=\frac{\Sigma_{sj}}{2\pi\sigma_{r j}\sigma_{\phi j}}
\exp\left\{-\frac{v_r^2}{2\sigma^2_{r j}}-
\frac{[v_\phi-v_c]^2}{2\sigma^2_{\phi j}}
\right\},
\label{distfun}
\end{eqnarray}
where $v_c$ is the circular velocity at current 
distance from the galactic center and $\sigma_{r j}$ and
$\sigma_{\phi j}$ are the velocity dispersions in $r$ and $\phi$
directions correspondingly. These velocity dispersions are related
by $\sigma_\phi^2/\sigma_r^2=4B^2/\kappa^2$,
where $B$ is Oort's B constant (Binney \& Tremaine 1987).
We will later use simply $\sigma$ to denote the velocity dispersion
in the r-direction $\sigma_r$.

For the description of collisional components we assume usual 
hydrodynamical  description with isotropic pressure.
To describe  the star-like components we use a different approach 
which is  rooted in the kinetic treatment of the 
collisionless systems as described elsewhere (Toomre 1964; Binney \& 
Tremaine 1987). We will see in \S \ref{2-fluid} that this accurate 
treatment shows that fluid approach really does 
quite a good job in describing the stability of the 
collisionless systems with multiple 
components as it does in one-fluid case, though some quantitative 
differences exist.

Equations governing the motion of 
the gas components in our coordinates are 
Euler's and continuity equations:
\begin{eqnarray}
\frac{\partial {\bf v}_i}{\partial t}+{\bf (v_{\it i} \nabla)v_{\it i}}=
-\frac{1}{\Sigma_i}{\bf \nabla}P_i -{\bf \nabla}\Phi, \label{mot}\\
\frac{\partial \Sigma_i}{\partial t}+
{\bf \nabla (}\Sigma_i{\bf v_{\it i})}=0\label{cont},
\end{eqnarray}
for $i$-th component.

We assume that pressure of each component
$P_i=K\Sigma_i^{\gamma}$ and
introduce specific enthalpy 
\begin{equation}
h_i=\frac{\gamma}{\gamma-1}K\Sigma_i^{\gamma-1}.
\end{equation}

We linearize equations (\ref{mot}) and (\ref{cont}) by assuming that
 $v_{r i}=u_i, v_{\phi i}=v_c+v_i, h_i=h_{0 i}+h_{1 i},
\Sigma_i=\Sigma_{0 i}+\Sigma_{1 i},$ and
$\Phi=\Phi_0+\Phi_1$.

Then equations (\ref{mot}) and (\ref{cont}) reduce in the first order to
(Binney \& Tremaine 1987)
\begin{eqnarray}
\frac{\partial  u_i}{\partial t}+\Omega
\frac{\partial u_i}{\partial \phi}-2\Omega v_i=
-\frac{\partial}{\partial r} (\Phi_1+h_{1 i}), \label{linmotr}\\
\frac{\partial  v_i}{\partial t}+\Omega
\frac{\partial v_i}{\partial \phi}-2 B u_i=
-\frac{1}{r}\frac{\partial}{\partial \phi} (\Phi_1+h_{1 i}),
 \label{linmotphi}\\
\frac{\partial \Sigma_{1 i}}{\partial t}+
\frac{1}{r}\frac{\partial}{\partial r}(r\Sigma_{i 0}u_i)+\Omega
\frac{\partial \Sigma_i}{\partial \phi}+\frac{\Sigma_{0 i}}{r}
\frac{\partial v_i}{\partial \phi}
=0,\label{lincont}
\end{eqnarray}
where
\begin{equation}
B=-\frac{1}{2}\left[\Omega+\frac{\partial (\Omega r)}{\partial r}\right]
\end{equation}
is the Oort's $B$ constant. 

Since here we are interested in axisymmetric
perturbations only it is  always supposed
that $\partial/\partial\phi=0$ in our analysis. 
We assume that all the first order dependent quantities vary like 
$\exp[i(kr-\omega t)]$, where $\omega$ is the angular frequency and
$k=2\pi/\lambda$ is the wavenumber. For the local analysis 
we employ the WKB approximation (or tight-winding approximation)
which requires that $kr\gg1$ and allows us to neglect terms
proportional to $1/r$ compared to the terms proportional to $k$.
With all these simplifications equations (\ref{linmotr})-(\ref{lincont})
reduce to
\begin{eqnarray}
-i \omega u_i-2\Omega v_i=-i k (\Phi_1+h_{1 i}),\label{rlin}\\
-i \omega v_i-2 B u_i=0,\label{philin}\\
-i \omega \Sigma_{1 i}+i k \Sigma_{0 i}u_i=0,\label{conlin} 
\end{eqnarray}
for $i$-th gas component.

These equations have to be supplemented with 
Poisson equation
\begin{equation}
\nabla^2 \Phi = 4\pi G \left(\sum\limits_{i=1}^{n_g}\Sigma_{1 i}+
\sum\limits_{j=1}^{n_s}\Sigma_j\right) \delta(z),
\end{equation}
where $\delta(z)$ is the Dirac delta function  arising 
from our assumption of infinitely thin disk. 
With $\Sigma_{1 i,j}$ and $\Phi_1$ being proportional 
to $\exp[i(kr-\omega t)]$
the relation between the perturbations of the potential and 
surface densities becomes
(Toomre 1964) 
\begin{equation}
\Phi_1= - \frac{2 \pi G}{k}\left(\sum\limits_{i=1}^{n_g}\Sigma_{1 i}+
 \sum\limits_{j=1}^{n_s}\Sigma_{1 j}\right).
\label{pot}
\end{equation}

The sound speed for each gas component is defined as 
$c_j^2=d P_j/d\Sigma_{0 j}$,
and in this case the perturbation of enthalpy reduces to
\begin{equation}
h_{1 j}=c^2_j\frac{\Sigma_{1 j}}{\Sigma_{0 j}}.
\label{enth}
\end{equation}

\section{Dispersion relation and stability criterion}\label{disp-rel}

Equations (\ref{rlin})-(\ref{conlin}), (\ref{pot}), and (\ref{enth})
form a closed set of linear equations and we now solve them to get 
the dispersion relation. 

From equations (\ref{rlin}) and (\ref{philin}) we can relate
perturbation of the radial velocity of each component $u_i$ to
the potential perturbation $\Phi_1$ using equation (\ref{enth}):
\begin{equation}
u_i=-\frac{\omega k}{\Delta}\left(\Phi_1+c_j^2\frac{\Sigma_{1 j}}
{\Sigma_{0 j}}\right),\label{rvel}
\end{equation}
where
\begin{equation}
\Delta=\kappa^2-\omega^2,
\end{equation}
and $\kappa^2=-4\Omega B$ is the epicyclic frequency.

Now we are in position to treat the star-like components. Since they are 
 supposed 
to be collisionless they cannot create any pressure and it means that 
for $j$-th stellar component sound velocity $c_j=0$. But because
of the 
 epicyclic motion at any given point in the disk
 there are stars from different parts
of the perturbed structure  and it leads to
important cancellation effects. 
They were first  calculated  by Toomre (1964)
and it can be shown that each stellar component is described by the Jeans
equations which are pretty similar to the usual hydrodynamic equations
with important difference being that instead of equation (\ref{rvel})
we have 
\begin{equation}
u_k=-\frac{\omega k}{\Delta} \Phi_1 
{\cal F}\left(\frac{\omega}{\kappa},\frac{k^2 \sigma_j^2}{\kappa^2}\right),
\label{rstar}
\end{equation}
where the reduction factor $F$ is given by the expression (Binney \& Tremaine
1987)
\begin{eqnarray}
{\cal F}(s,\chi)=\frac{2}{\chi}(1-s^2)e^{-\chi}\sum\limits^{\infty}_{n=1}
\frac{I_n(\chi)}{1-s^2/n^2},\label{redfac}
\end{eqnarray}
and $I_n$ are the Bessel functions of order $n$.

Now, from (\ref{conlin}), (\ref{rvel}) and (\ref{rstar}) we can
eliminate $u_i$ to get
\begin{equation}
\Sigma_{1 i}=-\frac{k^2 \Sigma_{0 i}}{\Delta+k^2 c_i^2} \Phi_1,
\label{gsig}
\end{equation}
 for $i$-th gas component and
\begin{equation}
\Sigma_{1 j}=-\frac{k^2 \Sigma_{0 j}}{\Delta}\Phi_1 {\cal F}_j,
\label{ssig}
\end{equation}
for $j$-th stellar component and ${\cal F}_j={\cal F}(\omega/\kappa,
k^2 \sigma_j^2/\kappa^2)$.

We can then substitute (\ref{ssig}) and (\ref{gsig}) 
into (\ref{pot}) to obtain finally
the desired dispersion relation:
\begin{eqnarray}
2\pi G k \sum\limits_{i=1}^{n_g}\frac{\Sigma_{0 i}}{\kappa^2+k^2 c_j^2
-\omega^2} + 2\pi G k \sum\limits_{j=1}^{n_s}\frac{\Sigma_{0 j} 
{\cal F}_j}{\kappa^2
-\omega^2}=1.
\label{disprel}
\end{eqnarray}

If only $2$ fluid components are present in the disk this dispersion relation
clearly reduces to the one derived by JS.
If one of the components is 
collisionless the dispersion relation is different and it was first 
considered by Romeo (1992). 
We discuss the difference between the two cases in \S 
\ref{2-fluid}.

The multicomponent disk is unstable if $\omega^2<0$ because in this case
oscillatory behavior of the density waves changes to an exponential growth.
The dispersion relation (\ref{disprel}) has an infinite number of solutions
with $\omega^2(k)>0$ for all $k$,
which are clearly stable. But there is also
one mode of oscillations in the system
 which has a single solution with $-\infty<\omega^2/\kappa^2<1$
and this mode could be unstable.

To find the condition for instability we note that as $\omega^2 \to -\infty$
at a fixed wavenumber $k$
LHS of the equation (\ref{disprel}) tends to zero, and
it monotonically increases to $+\infty$ as $\omega^2 \to \kappa^2$. 
This property of a {\it steady} growth is important because if
(\ref{disprel}) has no solution for negative $\omega^2$ it
implies that the LHS of (\ref{disprel}) at $\omega^2=0$ must be less than $1$
and vice versa. So the axisymmetric 
instability arises in multicomponent disk if and only if
\begin{eqnarray}
2\pi G k \sum\limits_{i=1}^{n_g}\frac{\Sigma_{0 i}}{\kappa^2+k^2 c_j^2}
 + 2\frac{\pi G k}{\kappa^2} \sum\limits_{j=1}^{n_s}
\Sigma_{0 j} \Psi_j > 1,
\label{stabil}
\end{eqnarray}
where 
\begin{eqnarray}
\Psi_j={\cal F}\left(0,
\frac{k^2 \sigma_j^2}{\kappa^2}\right)=\frac{2\kappa^2}{ k^2 \sigma_j^2}
e^{-k^2 \sigma_j^2/\kappa^2}\sum\limits^{\infty}_{n=1}
I_n\left(
\frac{k^2 \sigma_j^2}{\kappa^2}\right).
\label{psi}
\end{eqnarray}
Since
\begin{equation}
\sum\limits^{\infty}_{n=1}
I_n(\chi)=\frac{1}{2}\left[e^\chi-I_0(\chi)\right],
\end{equation}
(Abramowitz \& Stegun 1964) we can rewrite (\ref{psi}) as
\begin{eqnarray}
\Psi_j=\frac{1-e^{-\chi_j}I_0(\chi_j)}{\chi_j}=
\frac{1-e^{-k^2 \sigma_j^2/\kappa^2}I_0(k^2 \sigma_j^2/\kappa^2)}
{k^2 \sigma_j^2/\kappa^2}.
\label{psi1}
\end{eqnarray}

Suppose again that 
 there are only two fluid (collisional) components in the disk then the
instability condition (\ref{stabil})
reduces to the one derived in JS as it should.

\section{Realistic two-fluid case}\label{2-fluid}

Now we consider the stability in the 
case when we have only two fluids present in the disk in more details.
Our main purpose here is to see how
the fact that one of the components is in reality  collisionless
changes the overall dynamics of the system from the case considered
by JS. 

In the case of two components, one stellar and one gaseous,  we assume
 $\Sigma_g$ and $\Sigma_s$ to be the unperturbed gas and stellar 
surface densities, $c_g$ the gas sound speed, and
$\sigma_s$  the stellar velocity dispersion in $r$-direction and 
 introduce the following dimensionless quantities
\begin{eqnarray}
Q_s=\frac{\kappa \sigma_s}{\pi G \Sigma_s},~~
Q_g=\frac{\kappa c_g}{\pi G \Sigma_g},\\
q=k\sigma_s/\kappa, ~~ R=c_g/\sigma_s.
\end{eqnarray}

Here $Q_g$ is a Toomre's $Q$ parameter for gas, while $Q_s$ is different
from the Toomre's $Q$ parameter for the collisionless system, which is
given by $\kappa \sigma_s/(3.36 G \Sigma_s)$ (Toomre 1964). It means
that with our definition of $Q_s$ the 
one--component stellar disk is gravitationally unstable 
when $Q_s<3.36/\pi=1.07$. 

With these definitions our instability condition (\ref{2fl}) reduces to
\begin{eqnarray}
\frac{2}{Q_s}
\frac{1}{q}\left[1-e^{-q^2} I_0(q^2)\right]
+\frac{2}{Q_g}R\frac{q}{1+q^2 R^2}>1.
\label{dimsg}
\end{eqnarray}

If one follows the Jog \& Solomon approach and treats stellar component
as a fluid with sound speed equal to $\sigma_s$, one gets
 the following instability condition  in terms of
our dimensionless variables
\begin{eqnarray}
\frac{2}{Q_s}\frac{q}{1+q^2}
+\frac{2}{Q_g}R\frac{q}{1+q^2 R^2}>1.
\label{dimgg}
\end{eqnarray}

\begin{figure}
\vspace{16.cm}
\includegraphics{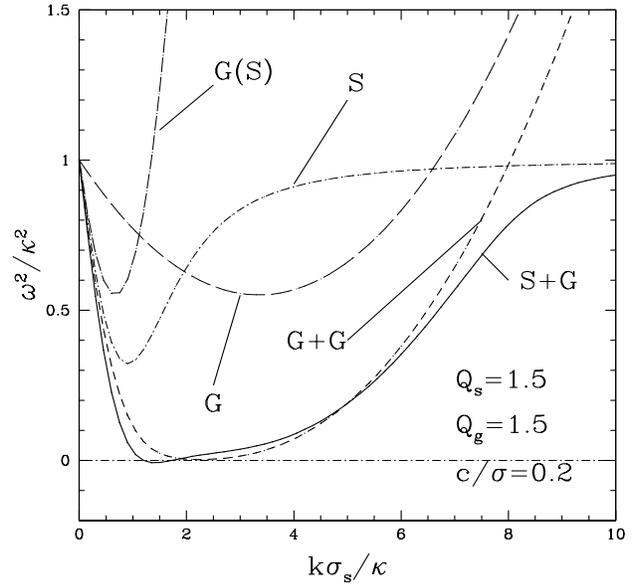}
\includegraphics{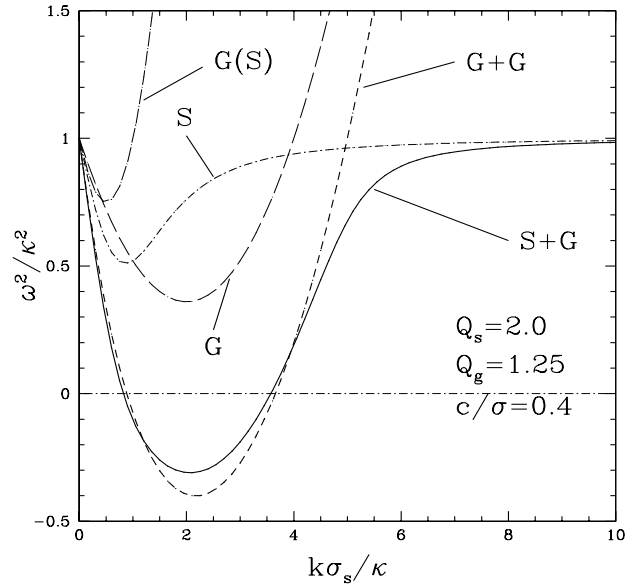}
\caption{Plots showing the 
dependence of the normalized square of the frequency $\omega^2/\kappa^2$ 
of the two-component disk 
$\omega^2$ upon the dimensionless wavenumber of the perturbation
 $k\sigma/\kappa$. Two
examples are shown: (top panel) $Q_g=Q_s=1.5$, $\Sigma_g/\Sigma_s=0.2$, 
and $c/\sigma=0.2$, and
(bottom panel) $Q_g=1.25, Q_s=2.0$, 
$\Sigma_g/\Sigma_s=0.65$,  and $c/\sigma=0.4$.
Labels near different curves mean: G$+$G - dispersion relation for
two fluid components, S$+$G - dispersion relation for stellar-fluid disk,
G - for cold gas only, S - for stars only, and G(S) - for gas
with the sound speed equal to the radial velocity dispersion of stars.
}
\label{omega_2}
\end{figure}

In Figure \ref{omega_2} we show the dependences of the $\omega^2$
upon $k$ for some particular choices  of parameters and
compare the curves in the case of gas-gas mix considered
 by Jog \& Solomon (labeled as G$+$G)
and star-gas mix (labeled as S$+$G), calculated using dispersion relation
(\ref{disprel}) for the case of two-component disk.
If one adopts $\kappa=36$ km s$^{-1}$ kpc$^{-1}$ to characterize the disk 
rotation and sound speed of the gas $c=5$ km s$^{-1}$, then the 
model corresponding to the Figure \ref{omega_2}a has
$\Sigma_s = 45~ M_{\odot}$ pc$^{-2}$, $\Sigma_g = 9 ~M_{\odot}$ pc$^{-2}$, 
and $\sigma=25$ km s$^{-1}$, which is a good representation
of the typical spiral galaxy like our own. 

Second plate of the same Figure 
correspond to the model with $\Sigma_s=17~ M_{\odot}$ pc$^{-2}$,
$\Sigma_g=11~ M_{\odot}$ pc$^{-2}$, and  
$\sigma=12.5$ km s$^{-1}$ for the same
choice of $\kappa$ and $c$. This might be thought of as a
central part  of a relatively young spiral and it turns out 
to be unstable with both instability criteria.

One can immediately see that in general
the difference 
between the curves given by conditions (\ref{dimgg}) and (\ref{dimsg}) 
is small for 
low enough wavenumbers as it should be because pressure 
forces are negligible in this regime and it is only the epicyclic 
motions and the self-gravity which determine the dynamics of the disk. 
However, as 
$k$ grows, in gas-gas case the frequency describing the oscillations of the 
whole system grows due to the pressure forces in the gas, while in the 
star-gas case this does not happen. Despite the presence of the
gas in the system it is really stars which determine the 
natural frequency of oscillations of the system in this case 
and as $k\to\infty$ frequency becomes constant -- $\omega \to \kappa$ --
and it presents a dramatic difference in the disk response 
to the axisymmetric density perturbations at small
wavelengths compared to the fluid-fluid case.

\begin{figure}
\vspace{7.5cm}
\includegraphics{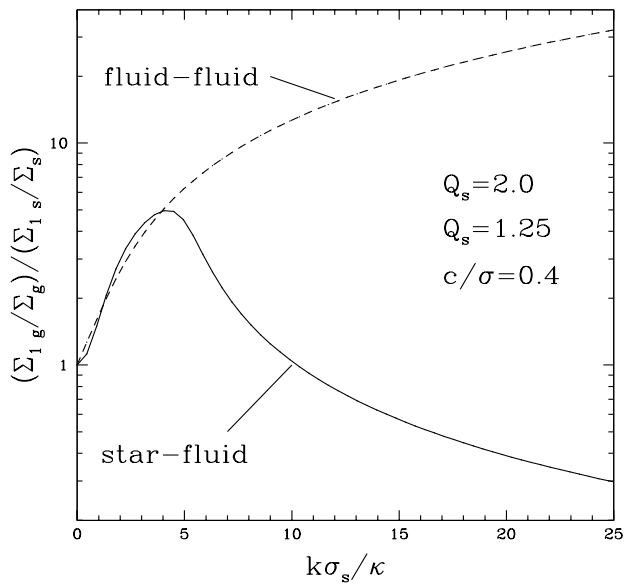}
\caption{
Plot showing ratio of the relative perturbations of
gas and stellar surface densities for two cases:
when stars are treated as fluid (dashed line) and when they are
considered as collisionless  (solid line). Disk parameters here are
those used in Figure \ref{omega_2} (lower panel). The distinction
between two approaches is obvious: as $k$ grows the  amplitude of density
perturbations in the gas relative to the density
perturbations in the stars
increases in the former case and decreases in the latter.
}
\label{ratiosig}
\end{figure}

In the two-fluid case JS found that 
the ratio of the perturbation of the gas surface density 
to that of the stellar surface density is always a growing function of 
wavenumber $k$ and for large $k$ this ratio becomes
very large. This is not the case when one of the disk components is
collisionless. Indeed, in this case $\omega\to\kappa$ for large $k$. 
Using (\ref{gsig}) and (\ref{ssig}) we get
\begin{eqnarray}
\frac{\Sigma_{1 g}}{\Sigma_{1 s}}\to
\frac{\Sigma_{g}}{\Sigma_{s}}\frac{\sigma^2}{c^2}
\left[2 e^{-q^2}\sum\limits^{\infty}_{n=1}
\frac{I_n(q^2)}{1-\omega^2/(\kappa^2 n^2)}\right]^{-1}
\to 0,
\label{r}
\end{eqnarray}
because the sum in (\ref{r}) diverges as $\omega\to\kappa$.
So,  in realistic gas-stellar case this ratio goes to $0$ rather
than increases. 
It is illustrated in Figure \ref{ratiosig} where
we have shown the ratio of relative perturbed surface densities
 for two cases: fluid-fluid and fluid-stellar
with the same set of parameters. 
This decrease in the gas density perturbation with growing $k$
leads to the dominant role of stellar component in the dynamics 
of the whole system for large $k$, which is different from the 
JS case.

The relative contribution to the instability per
unit surface density of the component considered by JS is given by the 
ratio of the corresponding terms in  the LHS of (\ref{dimgg})
\begin{eqnarray}
\gamma=\frac{\Sigma_g}{\Sigma_s}\frac{1+q^2}{1+q^2 R^2}
\end{eqnarray}
and is always larger than $\Sigma_g/\Sigma_s$ for $R<1$. In the  star-fluid
case one gets from (\ref{dimsg}) that
\begin{eqnarray}
\gamma=\frac{\Sigma_g}{\Sigma_s}\frac{q^2}{(1+q^2 R^2)
\left(1-\exp(-q^2)I_0(q^2)\right)}.
\end{eqnarray}
Careful study shows that for 
$\sqrt{3}/2<R<1$ there is always a range of small $q$ (and $k$) in which
$\gamma<\Sigma_g/\Sigma_s$. Of course, as $q$ grows $\gamma$ becomes
greater than $\Sigma_g/\Sigma_s$ for $R<1$ 
($\gamma\to R^{-2} \Sigma_g/\Sigma_s$ as $q\to\infty$ ), 
but for $q\sim 1$ the relative
contribution to the instability per unit surface density  is greater
for stars if $\sqrt{3}/2<R<1$. Even in this case though gas contributes
 only several per cent smaller
than the stars
 and  in most astrophysically interesting cases
 cold material still has a 
serious impact on the stability of the system.

\begin{figure}
\vspace{7.5cm}
\includegraphics{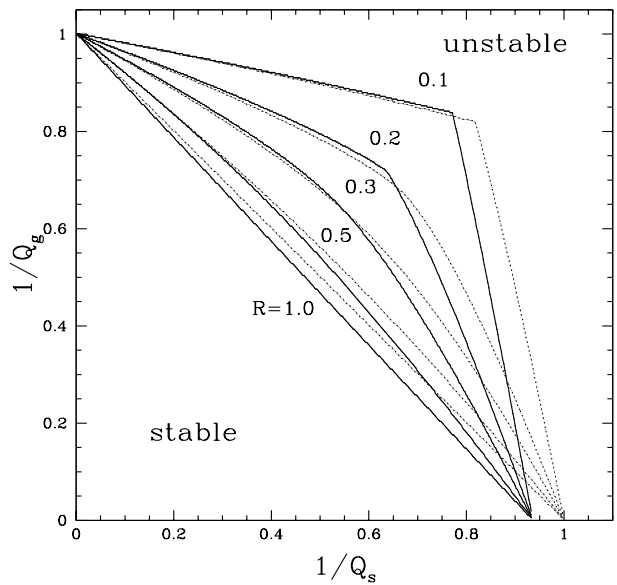}
\caption{
Plot showing stable and unstable models of the two-fluid disk.
Thick solid lines represent the marginal stability curves according
to the star-fluid instability criterion (\ref{dimsg}) while thin dotted lines
correspond to the Jog \& Solomon (1984a) fluid-fluid criterion (\ref{dimgg}).
Each curve is labeled with corresponding ratio of the gas sound
speed to the stellar velocity dispersion in $r$-direction $R=c/\sigma$.
The part of the parameter space bounded by each marginal stability 
curve is stable for a corresponding $R$.
One can see that the difference between two approaches -- star-fluid 
and fluid-fluid 
is most pronounced at $1/Q_s$ near $1$ and $1/Q_G$ near $0$ 
where the kinetic and fluid approaches
differ the most due to the weak influence of the gas  component on the
overall disk dynamics.}
\label{Figcompare}
\end{figure}

Each of the criteria (\ref{dimsg}) and (\ref{dimgg}) produces some 
region in parameter space of the disk models in which they are stable
against local axisymmetric gravitational perturbations.
Unfortunately, it does not seem trivial to construct an effective 
analytical way of defining some effective value $Q_{eff}$
as a function of $Q_g$, $Q_s$ and other disk parameters so 
that the system is stable when $Q_{eff}>1$, as it was done for
the fluid-fluid case by Elmegreen (1995). 
Instead we follow the approach of Jog (1996) 
and study the problem seminumerically. 
We parametrize our models here by $1/Q_s, 1/Q_g$ and $R$.
In Figure \ref{Figcompare} we compare the stable regions produced 
by each of the instability criteria.

In the JS case the marginal stability curves given by 
conditions  $\partial\omega^2/\partial k=0$ and $\omega^2=0$ are
symmetric with respect to the line $Q_s=Q_g$. Indeed, if these
conditions are fulfilled
and an inequality in (\ref{dimgg}) changes to equality 
 for $Q_s=Q_1, Q_g=Q_2$ and $q=q_{crit}$, 
then one can easily check that this is also true for $Q_s=Q_2,
Q_g=Q_1$ and $q=q_{crit}/R$ provided that
 the instability condition is given by
(\ref{dimgg}). Of course, this is not the case for the star-gas disk
because any such symmetry is absent in the relation (\ref{dimsg}).

As we increase $R$ from $0$ to $1$ the region of the parameter space occupied
by stable models shrinks until $R=1$. 
The further increase of $R$ beyond $1$ causes reexpansion of the region
occupied by stable models. In fact in the fluid-fluid case the marginal curve
corresponding to some particular $R=R_0$ coincides 
with the curve corresponding
to $R=1/R_0$ which can be directly checked using (\ref{dimgg}). 
But these models  
are likely to be uncommon since they have $c>\sigma$ which 
seems to be unusual in real galaxies.

The stable region in the star-gas case is in general smaller
than that in the gas-gas case. The difference is especially noticeable 
at  $1/Q_s\approx 1$ and low $1/Q_g$, when the gas influence on the 
dynamics of the system is 
smallest and the stability condition is close to the one-fluid stellar 
stability
criterion which is somewhat different from 
the fluid case. 
Nevertheless, the difference between two cases is quite small for most of
the parameters (especially for small $Q_s$) 
and one can usually use the JS stability 
criterion in these cases.

\section{Application to the Solar neighborhood}\label{appl}

In this section we  apply the results derived in  \S \ref{disp-rel}
to the neighborhood of the Sun in the Galaxy. 

In doing this one should realize that all the stars and gas in the Galaxy
cannot be simply put into two distinct groups with some well defined 
velocity dispersion for stars and sound speed for gas. The reason for that
 is that the stars of different ages have different velocity 
dispersions - the older the star the larger its random motion. This 
random heating of stellar population
is produced by the scattering of the stars by the giant molecular clouds
(Spitzer \& Schwarzschild 1951, 1953)
and/or transient spiral density waves in the Galaxy (Barbanis \& Woltjer 1967;
Carlberg \& Sellwood 1985). 

Gas in its turn intrinsically has a multicomponent nature caused by the
constant energy input from supernovae explosions and
various cooling and heating processes determining its thermal equilibrium
(McKee \& Ostriker 1977; Kulkarni \& Heiles 1987).
Recent studies have enabled us to distinguish 
5 phases of the ISM: molecular gas in the form of the clouds 
(Scoville \& Sanders 1987), 
cold neutral medium (CM) also in the form of clouds, and 3 more or less
uniformly distributed gaseous components: warm neutral medium (WNM),
warm ionized medium (WIM) (Kulkarni \& Heiles 1987), 
and hot component (Savage 1987).  They
have different sound speeds and surface densities which makes one 
treat them separately. 

Unperturbed gravitational field of our Galaxy is not axisymmetric
and it limits our analysis to some extent. It was also shown that
the stellar distribution function can not always be represented
by formula (\ref{distfun}) because nonaxisymmetries of the Galactic
gravitational field cause vertex deviations of the velocity
ellipsoid (Dehnen \& Binney 1998). Young stars of O and B types
are likely not to have Schwarzschild distribution because
they have had no time to be sufficiently scattered by the giant
molecular clouds or transient spiral arms, but they probably are not
important mass contributors in the local part of the Galaxy.
The local approximation itself maybe questionable because
most unstable waves have $\lambda$ of the order of several kpc. 

Other possible complications in real galactic disks involve the presence of 
cosmic rays and magnetic fields which could  influence the 
gas dynamics (Elmegreen 1987).
In this paper we are primarily interested in the purely gravitational
aspects of the disk instability and for this reason we neglect
them at all, though it makes our consideration  less 
realistic when applied to the galaxies.
For these reasons our analysis of the  stability
of the Solar neighborhood should be considered only as mostly 
illustrative though bearing sufficient resemblance to reality. 

Following Holmberg \& Flynn (2000) we split the galactic disk mass
between $13$ major parts: $4$ gaseous and $9$ stellar. We neglected
hot component of the gas because of its low number density, $n\sim 0.003$
cm$^{-3}$, high temperature, $T\sim 10^6$ K, (Savage 1987) 
and, consequently, large thickness which makes the reduction 
effects very important (Romeo 1992). 
For the same reason we neglect the stellar halo component.
Parameters of the gaseous components are taken from 
Kulkarni \& Heiles (1987)
 and Scoville \& Sanders (1987).

Surface densities of stellar  components are  taken from  
 Holmberg \& Flynn (2000). 
We got the radial velocity dispersions based on the recent 
data from $Hipparcos$ from 
Mignard (2000). 
Velocity dispersions of white and brown dwarfs have been 
chosen quite arbitrarily. All the parameters
assumed for the mass constituents are
listed in Table 1.

\begin{center}
\begin{tabular}{|l l l l|}
\hline
i & Component & $\Sigma_i$ & $\sigma_{r i}$ or $ c_s$ \\
  &           & ($M_{\odot}~{\rm pc}^{-2}$) & (${\rm km~s}^{-1}$) \\
\hline
$1$ &   ${\rm H}_2$ & $3.0$ & $4.0$\\ 
\hline
$2$ &   CM & $4.0$ & $6.9$\\ 
\hline
$3$ &   WNM & $4.0$ & $9.0$\\ 
\hline
$4$ &   WIM & $2.0$ & $9.0$\\ 
\hline
$5$ &   giants & $0.4$ & $26.0$\\ 
\hline
$6$ &  $M_V<2.5$ & $0.9$ & $17.0$\\ 
\hline
$7$ &  $2.5<M_V<3.0$ & $0.6$ & $20.0$\\ 
\hline
$8$ &  $3.0<M_V<4.0$ & $1.1$ & $22.5$\\ 
\hline
$9$ &  $4.0<M_V<5.0$ & $2.0$ & $26.0$\\ 
\hline
$10$ &  $5.0<M_V<8.0$ & $6.5$ & $30.5$\\ 
\hline
$11$ &  $M_V>8.0$ & $12.3$ & $32.5$\\ 
\hline
$12$ & white dwarfs & $4.4$ & $32.5$\\ 
\hline
$13$ & brown dwarfs & $6.2$ & $32.5$\\ 
\hline
\label{table2}
\end{tabular}
\end{center}

\begin{figure}
\vspace{7.5cm}
\includegraphics{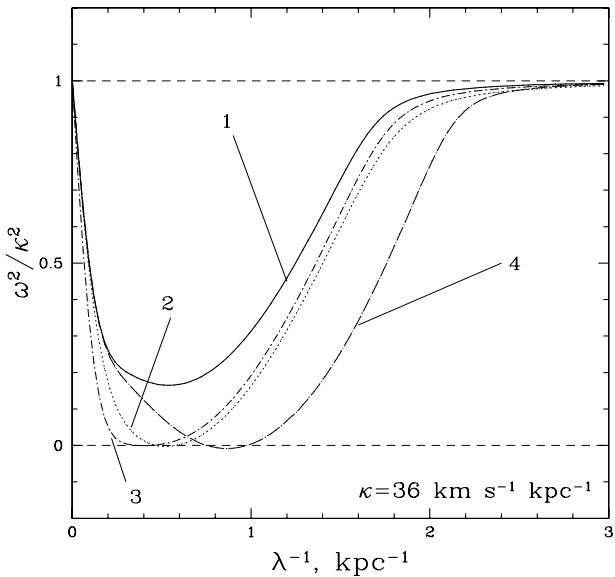}
\caption{
Plot showing the dependence of $\omega^2$ upon $\lambda^{-1}$ in the 
Solar neighborhood. Thick solid curve labeled $1$ corresponds to the model
with parameters given in Table 1. Other curves correspond 
to a different sets of parameters described in the text.
The model of the Solar neighborhood described in Table 1
is stable against local axisymmetric gravitational perturbations.
}
\label{solnei}
\end{figure}

In Figure \ref{solnei} we show the dependence of $\omega^2$ upon
the inverse radial wavelength $k/2\pi$. 
Epicyclic frequency $\kappa=36$ km~s$^{-1}$~
kpc$^{-1}$ is assumed throughout the calculation.
Curve labeled $1$ corresponds to the choice of parameters described
in the Table 1. One can see that for all radial wavelength 
$\omega^2$ is positive that is the whole disk system is stable against 
local axisymmetric perturbations. To check how rigorous this conclusion is
we varied some of the model parameters until the disk became unstable.

Data about the surface densities and velocity dispersions 
of the brown dwarfs (BDs) and white dwarfs (WDs) seem to 
be the most uncertain
among all the model parameters, so we tried to vary them first. 
Curve labeled $2$ has all the surface densities as listed in 
Table 1 but the velocity dispersion of WDs and
BDs was lowered to $\sigma_r=19.2$ km~s$^{-1}$. Only if
this population is so cold can it make the system unstable with
all other parameters being kept unchanged. It seems inevitable
 that real
WD and BD populations must be sufficiently hotter because only
young stars can have such a low velocity dispersion (Mignard 2000).

Curve labeled $3$ shows $\omega^2-\lambda^{-1}$ dependence for
the case when total surface density of WDs and BDs was raised from 
$10.6~ M_\odot$ pc$^{-2}$ to $26~ M_\odot$ pc$^{-2}$ keeping the
rest of model parameters unchanged. This leads to the neutral stability
of the system but such a surface density seems to be too 
large despite large uncertainties and claims of some authors that 
such extreme values of surface density could be common. 
For example, Festin (1998) found
high mass density of the WD, $\sim 2.6$ times larger than we assume here,
but his conclusions were based on a small sample of $7$ sources only.
Other authors (Ruiz \& Takamiya 1995; Oswalt et al. 1996) claim values
for WD mass density which are in agreement with what we take.
Even more uncertainty is involved in determining the density of the BD.
Different surveys quote values from $0.6$ to $4$ times what we assume 
in this research (Fuchs, Jahreiss, \& Flynn 1998). Recent data
(Reid et al. 1999) based on a large enough sample imply mass density of BD 
$0.005~M_\odot$ pc$^{-3}$
which is about $60\%$ of what we assume in our calculations.

Finally, the fourth curve shows the dispersion relation for the
disk with the lowered sound speed in some of the gas components:
in CM we set $c_s=5.0$ km~s$^{-1}$, in WNM $c_s=7.5$ km~s$^{-1}$,
and in WIM  $c_s=8.0$ km~s$^{-1}$, with all other parameters
unchanged. In this case disk becomes marginally unstable. 
Important thing to notice here is 
 that small variations in the gas sound speed can have 
stronger influence on the disk stability than large changes in the
stellar velocity dispersion, even though the surface density of
the gas is smaller than that of stars. This is a manifestation
of the crucial importance  of cold material
for the stability of the whole disk, which was first noted by 
Jog \& Solomon (1984a).
It is also easy to see that variations of the gas parameters produce
significant change of the most unstable wavelength compared to
the variations of the stellar parameters; it is reduced from
$\sim 2$ kpc to $\sim 1$ kpc by that small change in gas sound speeds.

The bottom line is that local Galactic disk seems to be stable against 
local gravitational axisymmetric perturbation even when allowance for
 a scant knowledge  of some of the Galactic parameters is made, which
confirms results of Jog \& Solomon (1984b) and Elmegreen (1995) for
two-fluid disks.

\section{Conclusions}\label{concl}

Due to the continuous interaction with the transient spiral
structure and giant molecular clouds stars diffuse in the velocity space
towards higher random velocities as their age increases and
it was confirmed observationally (Dehnen \& Binney 1998; Mignard 2000).
It raises a necessity of considering the dynamics of the Solar neighborhood
taking into account its complex multicomponent structure. In this paper
we studied the stability of such a system against gravitational axisymmetric
perturbations in the tight-winding  limit. It is possible to derive 
an analytic dispersion relation characterizing multicomponent thin
differentially rotating disk and study its stability. 

In doing so we distinguished between two types of the disk constituents:
stellar and gaseous. Stellar population is dynamically 
different from fluid because stars form collisionless system 
(Binney \& Tremaine 1987) while  gas must be treated as a fluid.  
We demonstrated that the difference in the results for stability 
produced  by two approaches is small 
for multicomponent disks in many astrophysically interesting
cases. Some disk models though could be sensitive to the choice of 
the stability condition and in that case one should use correct 
criterion given by the equation (\ref{stabil}).

We apply our results to  the stability
of the Solar neighborhood and confirm the conclusions of the 
previous two-fluid studies that local Galactic disk is stable against
axisymmetric perturbations in the WKB limit, even taking into account 
uncertainties associated with determining some of the disk parameters.

Keeping in mind  previous two-fluid results (JS)
it is not surprising that relatively small variations of the gaseous 
component parameters are very important for the overall disk 
stability. Indeed, Figure \ref{solnei} shows that decrease of the sound speed 
of the gas by about $1$ km s$^{-1}$ drives instability of the disk, 
while in the case of stellar component one needs to reduce its velocity
dispersion by $\sim 10$ km s$^{-1}$ to produce the same outcome.
Even though the mass of the gas in the disk is smaller that the stellar mass 
its small random motion makes it much more susceptible to its own
self-gravity than the hot stellar component. 

Our study of the stability of the Solar neighborhood neglects a lot
of physics such as magnetic fields or nonaxisymmetry of the 
Galactic gravitational field and their importance remains an open question. 
Measurement errors associated with 
determining some disk parameters  also limit the applicability
of multicomponent criterion 
because the larger the number of constituents the larger
errors get accumulated. Future interferometric missions
such as $GAIA$ and $SIM$ will probably solve this problem because
of their anticipated accuracy.
Nevertheless, even with all the simplifications and 
the observational uncertainties involved
it seems that the Solar neighborhood is
stable against purely gravitational
axisymmetric perturbations.

\section{Acknowledgements} 

It is my great pleasure to thank S.Tremaine for reading the manuscript
and making valuable suggestions and B.T.Draine for useful discussions.
Author
would also like to acknowledge the financial support of this work by
the Princeton University Science Fellowship.

\end{document}